\documentclass{ieeeaccess}

\usepackage{cite}
\usepackage{amsmath,amssymb,amsfonts}

\usepackage{graphicx}
\usepackage{textcomp}
\usepackage{verbatim}
\usepackage{url}
\usepackage{adjustbox}
\usepackage{multirow}
\usepackage{romannum}
\usepackage{csquotes}
\usepackage{dirtytalk}
\usepackage{subcaption}
\usepackage{dblfloatfix} 
\usepackage[pdftex,pdfpagelabels,bookmarks,hyperindex,hyperfigures]{hyperref}

\usepackage{cite}
\usepackage{algorithmic}
\usepackage{textcomp}
\usepackage{multirow}
\usepackage{latexsym}
\usepackage{graphicx}
\usepackage{amsfonts,amssymb,amsmath}
\usepackage{xcolor}
\usepackage{mathtools}
\usepackage{stmaryrd}
\usepackage{multirow}
\usepackage{booktabs}
\usepackage{adjustbox}
\usepackage{caption}
\usepackage{subcaption}
\usepackage{comment}
\usepackage[ruled,vlined]{algorithm2e}

\usepackage{lipsum}
    
\begin{document}

\history{Date of publication xxxx 00, 0000, date of current version xxxx 00, 0000.}
\doi{10.1109/ACCESS.2017.DOI}

\title{A Graph Neural Networks based Framework for Topology-Aware Proactive SLA Management in a Latency Critical NFV Application Use-case}
\author{\uppercase{Nikita Jalodia}\authorrefmark{1,2,3}, \IEEEmembership{Member, IEEE},
\uppercase{Mohit Taneja}\authorrefmark{1,4}, \IEEEmembership{Member, IEEE}, \uppercase{and Alan Davy}\authorrefmark{1,2,3}, \IEEEmembership{Senior Member, IEEE}}

\address[1]{Walton Institute for Information and Communication Systems Science, South East Technological University, Waterford, X91WR86, Ireland}

\address[2]{Department of Computing and Mathematics, School of Science and Computing, South East Technological University, Waterford, X91K0EK, Ireland}

\address[3]{CONNECT--- Centre for Future Networks and Communications, Dublin, Ireland}

\address[4]{Department of Accounting and Economics, School of Business, South East Technological University, Waterford, X91 TX03, Ireland}

\tfootnote{This work has emanated from research conducted with the financial support of Science Foundation Ireland (SFI) and is co-funded under the European Regional Development Fund under Grant Number 13/RC/2077.}

\markboth
{N. Jalodia \headeretal: Preparation of Papers for IEEE TRANSACTIONS and JOURNALS}
{N. Jalodia \headeretal: Preparation of Papers for IEEE TRANSACTIONS and JOURNALS}

\corresp{Corresponding author: Nikita Jalodia (e-mail: nikita.jalodia@waltoninstitute.ie, nikita.jalodia@postgrad.wit.ie).}

\begin{abstract}
Recent advancements in the rollout of 5G and 6G have led to the emergence of a new range of latency-critical applications delivered via a Network Function Virtualization (NFV) enabled paradigm of flexible and softwarized communication networks. Evolving verticals like telecommunications, smart grid, virtual reality (VR), industry 4.0, automated vehicles, etc. are driven by the vision of low latency and high reliability, and there is a wide gap to efficiently bridge the Quality of Service (QoS) constraints for both the service providers and the end-user. In this work, we look to tackle the over-provisioning of latency-critical services by proposing a proactive SLA management framework leveraging Graph Neural Networks (GNN) and Deep Reinforcement Learning (DRL) to balance the trade-off between efficiency and reliability. To summarize our key contributions: 1) we compose a graph-based spatio-temporal multivariate time-series forecasting model with multiple time-step predictions in a multi-output scenario, delivering 74.62\% improved performance over the established baseline state-of-art model on the use-case; and 2) we leverage realistic SLA definitions for the use-case to achieve a dynamic SLA-aware oversight for scaling policy management with DRL.
\end{abstract}


\maketitle

 \begin{figure*}[t]
	\centering
	\includegraphics[scale=0.6]{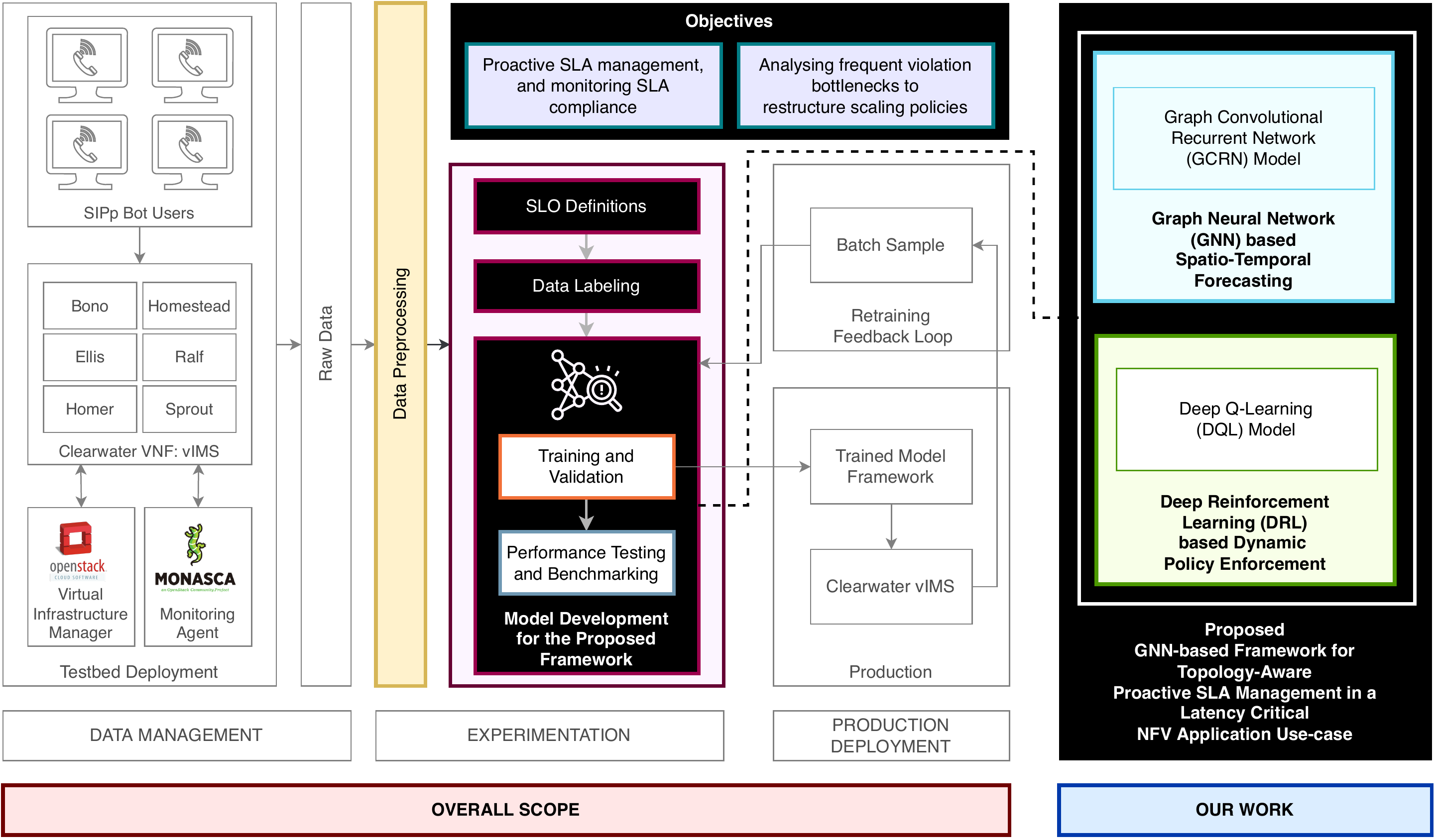}
	\caption{Overview of system architecture, objectives, and highlighted scope.}
	\label{fig:clearwater-testbed}
\end{figure*}

\section{Introduction} \label{sec:introduction} 
\PARstart{T}{he} next generation of communication networks continue to be driven by a fundamental restructuring in the way that the networks and services are deployed and delivered. 5G's usage scenario of ultra-reliable low-latency communications (URLLC) is further expected to extend in scope to a high-throughput ubiquitous global connectivity at scale, driving all major verticals towards a change \cite{jiang_road_2021}. Network programmability and softwarization are the key drivers of this shift that extends to the vision of 6G\cite{jiang_road_2021}, and are delivered via the concepts of Software Defined Networking (SDN) and Network Function Virtualization (NFV) \cite{mijumbi_network_2016}. These continue to play a pivotal role in the vision of 6G, and together form the backbone of flexible and intelligent networks. 

As Cloud infrastructure transitions to serve the next-generation requirements of upcoming softwarized application verticals \cite{mijumbi_network_2016}, this is met with evolutionary challenges of incremental reliability guarantees that are expected to out-do legacy systems with specialized infrastructure \cite{jiang_road_2021}. To this end, network slicing clusters applications with similar demands in an appropriate Cloud environment \cite{zhang_nfv_2021}, ensuring the placement of latency-critical URLLC applications in a high availability slice where resources are suitably provisioned to ensure reliability.

An SLA often invokes a trade-off between system usage and requirements, i.e., efficiency and reliability \cite{mijumbi_topology-aware_2017}. Over-provisioning is a common practice adopted by network and Cloud operators to match the high requirements for these latency-critical applications in a high availability network slice, but this is inefficient in the long run \cite{jalodia_deep_2019}. A transition towards complete softwarization of networks and services brings in the requirement to adopt more complex models to guarantee reliability and Quality of Service (QoS) \cite{cherrared_lumen_2018}.  A key aspect to driving such a change is in how the Cloud reacts to such a latency-critical demand, and in being precisely proactive over time \cite{jalodia_deep_2019, jalodia_deep_2021}.
 
In this work, we present a graph-based deep learning framework for dynamic and proactive SLA management in the use case of a latency-critical NFV application. An overview of the system and scope is provided in FIGURE \ref{fig:clearwater-testbed}.The key contributions are summarised as follows:
\begin{itemize}
    \item We work with a real-world deployment of a latency critical NFV application with two months' worth of raw network telemetry data sampled every 30 seconds, and use that as the basis for all our policy formation and learning models. 
    \item We compose a graph-based multivariate time-series forecasting model with multi-step predictions in a multi-output scenario, i.e. the model forecasts a sequential range of future values for multiple features in one go, thereby enabling us to track a realistic deployment setup. Further, we propose the suitability of a topology-aware Graph Neural Network (GNN) based model as opposed to traditional Recurrent Neural Network (RNN) methodologies when applied to such a task. Specifically, our implementation of a Gated Recurrent Unit (GRU) based Graph Convolutional Recurrent Network (GCRN) model demonstrates a 74.62\% improvement in performance over the established state-of-art model \cite{jalodia_residual_2022} on the use-case. 
    \item We leverage realistic Service Level Objective (SLO) definitions defined in our previous work \cite{jalodia_deep_2021} to compose a Q-learning based Deep Reinforcement Learning (DRL) model to achieve dynamic SLA-aware policy enforcement for such a latency critical use-case in an operational setting. 
\end{itemize}

To the best of our knowledge, this is the first approach in the area that proposes a GNN and DRL based framework for proactive SLA management, and widely outperforms the previously established residual connections based Long Short-Term Memory (Residual-LSTM) model benchmark \cite{jalodia_residual_2022} in the use-case's predictive objectives. \\

The rest of the paper has been structured as follows: \textsection \ref{related-work} elaborates on the background and presents the related work, \textsection \ref{defining-sla-slo} describes the Clearwater NFV application, and provides an overview of the SLA leveraged in the framework. \textsection \ref{proactive-sla-framework} describes the proposed framework in detail. Thereafter, \textsection \ref{experimental-setup} provides details of the experimental setup, \textsection \ref{results-discussion} discusses the results obtained through the proposed framework, and \textsection \ref{conclusion-futurework} presents the concluding remarks, with the scope of future work.


\section{Motivation and Related Work} \label{related-work}
A transition towards softwarized networks demands higher complexity in models to guarantee reliability and QoS \cite{cherrared_lumen_2018}, and improve policy-based QoS management \cite{binsahaq_survey_2019}. This is because of an impending evolution in not just the way networks are composed and managed, but also renewed application architectures \cite{cherrared_lumen_2018, cherrared_survey_2019}, corresponding QoS and SLA management techniques \cite{zheng_service_2021}, and optimization and automation to cope with the added complexity \cite{suomalainen_machine_2020}.

While service operators come up with new dynamic \cite{noauthor_dynamic_nodate} and predictive \cite{noauthor_predictive_nodate-1, noauthor_predictive_nodate, noauthor_using_nodate} scaling policies to match the demand facing the current generation of Cloud based application services, these are still a long way to go towards supporting latency-critical applications with high availability values \cite{jalodia_deep_2021}. Post appropriate provisioning following anticipated and identified traffic patterns, there is not a lot of work that directly addresses the remaining SLA bottlenecks from an application perspective. Automated SLA management for use-cases deployed on softwarized networks has been highlighted to be a critical requirement for next generation networks \cite{sun_sla-nfv_2016, yi_comprehensive_2018, kapassa_slas_2018, ben_yahia_cognitive_2017, bendriss_ai_2017, xli_automated_2021}. 

Much of the work done so far addresses QoS with characterizing and anticipating traffic patterns, anomaly detection \cite{hong_machine_2020}, and a combination of reactive and proactive scaling policies \cite{de_vleeschauwer_5growth_2021}, \cite{lee_deep_2020}, \cite{jiang_fast_2020}. Related work in the area explores anticipated network traffic based clustering and forecasting \cite{le_applying_2018, gebremariam_applications_2019}, machine learning based network traffic classification in NFV \cite{vergara-reyes_ip_2017, ilievski_efficiency_2020}, and related resource allocation \cite{jalodia_deep_2019, mijumbi_topology-aware_2017}. Our previous work \cite{jalodia_deep_2021} in the area proposes the use of multi-label classification methodology for a multi-output SLO violation prediction in NFV environments. From a feature forecasting perspective, authors in \cite{ferreira_forecasting_2021} present an overview of non-linear and linear forecasting methodologies to improve multi-slice resource management in 5G networks. LSTM based approaches \cite{bendrissForecastingAnticipatingSLO2017, abbasi_deep_2021, ferreira_forecasting_2021} have also been successfully leveraged for resource forecasting within communication networks. Our previous work in the area \cite{jalodia_residual_2022} benchmarks various deep learning based forecasting methodologies, and proposes a Residual LSTM based framework for proactive SLA management in rapid forecasting based resource monitoring of latency sensitive NFV applications. \\

However, networks can inherently be represented in a graph based format of nodes and edges, and so can the application structure and flow of data in the softwarized domain of SDN and NFV \cite{mijumbi_topology-aware_2017, jalodia_deep_2019}. In traditional Machine Learning (ML) approaches, these spatial inter-dependencies are removed in the pre-processing stage, and so the inherent effect of these topological dependencies are not a part of the learning and prediction stage \cite{scarselli_graph_2009}. Graph neural networks (GNNs) are a family of neural networks that deal with signals defined over graphs. Modern GNNs are categorized into four groups: recurrent GNNs, convolutional GNNs, graph autoencoders, and spatial–temporal GNNs \cite{wu_comprehensive_2021}. Further, in contrast to conventional Machine Learning (ML) approaches, a GNN based approach is capable of producing accurate predictions even when the underlying topology is changed from what the model was trained on \cite{ferriol-galmes_scaling_2021}.

Within the wider domain of networks, GNN has been successfully used in problem areas addressing networking performance and generalisation to larger networks \cite{ferriol-galmes_scaling_2021}, radio resource management \cite{shen_graph_2020}, etc. Such methodologies have also been successfully used in the area of SDN, addressing connection management \cite{orhan_connection_2021}, energy-efficient VNF deployment \cite{qi_energy-efficient_2021}, spatio-temporal link state prediction \cite{yeom_graph_2021}, detecting and mitigating data plane attacks \cite{cao_detecting_2021}, predicting the optimal path for Service Function Chain (SFC) deployment and packet-level traffic steering \cite{rafiq_service_2020}, etc. Within the domain of NFV, GNN based methodologies have proven successful in problem areas addressing network slicing management \cite{wang_graph_2020}, VNF deployment prediction \cite{kim_graph_2020}, VNF resource prediction \cite{mijumbi_topology-aware_2017}, finding optimal SFC path \cite{heo_graph_2020}, etc. 

Since GNN successfully extracts and models spatial features and topological dependencies, there is an intrinsic potential to use this in conjunction with a reinforcement learning (RL) based approach \cite{jalodia_deep_2019}. Authors in \cite{zhu_network_2021} have used graph convolutional network (GCN) based GNN with actor-critic based DRL towards network planning. The combination of GNN and DRL has also been successfully leveraged to tackle power grid management \cite{yoon_winning_2021}. Within the domain of SDN, message passing GNN with deep Q-learning has been used towards connection management \cite{orhan_connection_2021}, and routing optimization \cite{swaminathan_graphnet_2021}. Authors in \cite{qi_energy-efficient_2021} have used graph convolutional network (GCN) based GNN with double deep Q-network (DDQN) towards energy-efficient VNF deployment. Within the subject area of NFV, graph network based GNN and DRL have been effectively used towards NFV flow migration \cite{sun_deepmigration_2020}, \cite{sun_efficient_2020}, and optimal VNF placement \cite{sun_combining_2021}. Graph convolutional network based GNN, and DRL have also been leveraged to address the problem areas involving VNF forwarding graph placement \cite{xie_virtualized_2021}, and for virtual network embedding \cite{rkhami_use_2020}, \cite{yan_automatic_2020}.

This publication is an extension of our previous work in the area--- our first publication \cite{jalodia_deep_2021} in the series fulfilled the gap of a lack of realistic SLO definitions used within proposed models in research. We also proposed \cite{jalodia_deep_2021} the use of multi-label classification methodology for a multi-output SLO violation prediction in NFV environments. Next, our work in \cite{jalodia_residual_2022} benchmarked various deep learning based forecasting methodologies, and proposed a Residual LSTM based framework for proactive SLA management in rapid forecasting based resource monitoring of latency sensitive NFV applications. To the best of our knowledge, the proposition in this work is the first approach in the area that proposes a spatio-temporal Graph Convolutional Recurrent Network (GCRN) model for the use-case, and benchmarks its performance against conventional deep learning models that have demonstrated favorable performance on the use-case in the past. We also use realistic SLO definitions to propose an SLA-aware deep Q-learning based DRL model, packaged together in a framework that delivers topology-aware proactive SLA management in a latency-critical NFV application.


\section{Service Level Agreements (SLA)} \label{defining-sla-slo}
An SLA is a qualitative measure that binds the service provider and facilitator into a formally agreed contract ensuring QoS for the end user. This is realised quantitatively on a set of low level metrics delivered through SLOs and Service Level Indicators (SLIs). The SLIs are quantitative measures that build upon raw system metrics such as those presented in Table \ref{T1}, which further feed into the SLOs as a definitive target range or threshold towards the deliverance of an SLA. 
\begin{equation}
    SLI \leq target \ threshold
    \label{eq1}
\end{equation}
\begin{equation}
    lower bound \leq SLI \leq upper bound
    \label{eq2}
\end{equation}
The breach of an SLA implies an explicit consequence, often financial; while the SLOs and SLIs are typically measurable indicators that define the policy of tolerance with measurable service characteristics \cite{beyer_site_2016}.

\subsection{Project Clearwater Cloud IMS}
\begin{figure}[t]
	\centering
	\includegraphics[width=\linewidth]{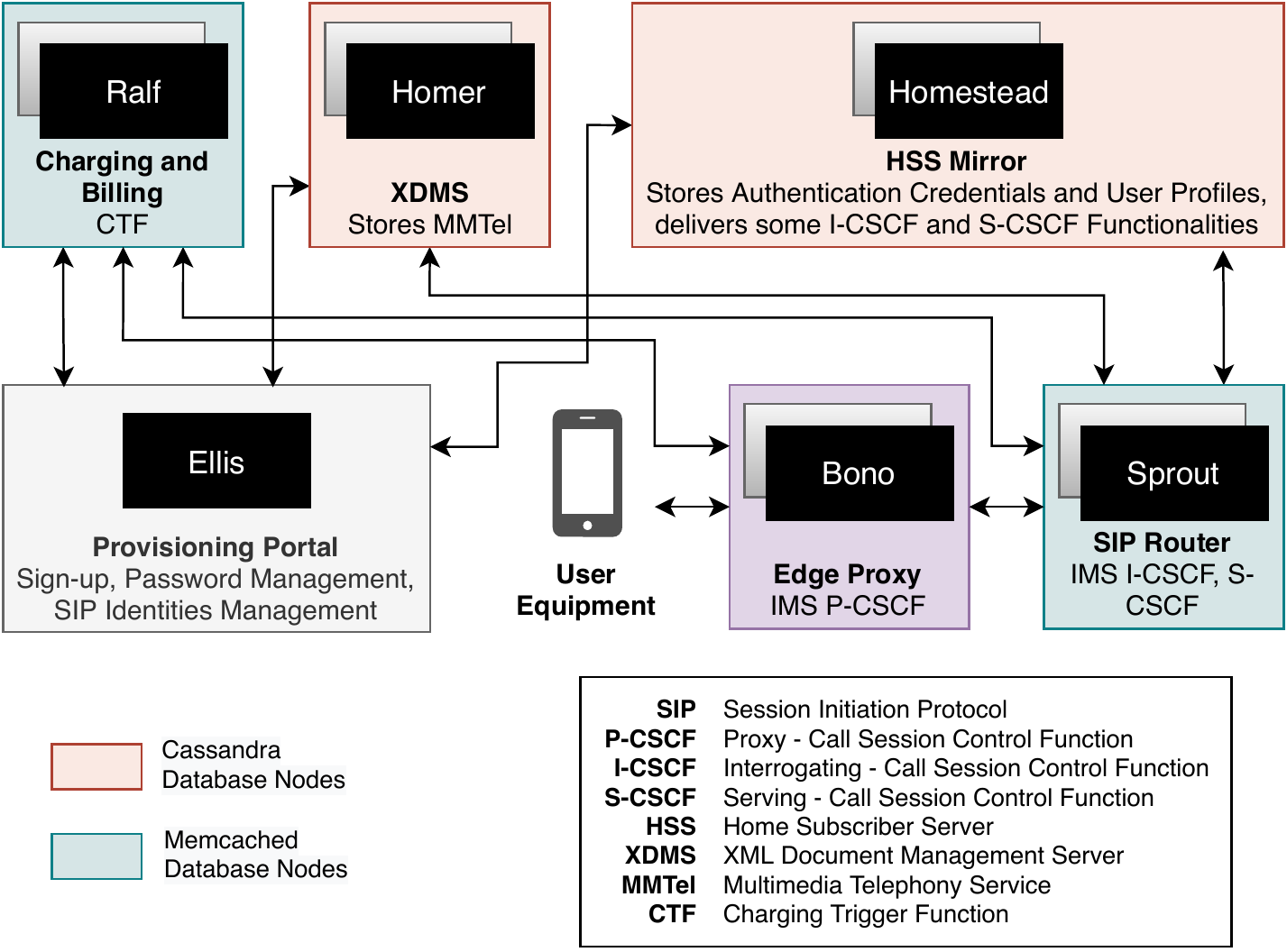}
	\caption{Clearwater vIMS architecture, depicting the various VNFCs and their high-level functionalities.}
	\label{fig:clearwater}
\end{figure}

In our work, we use the open-source Project Clearwater\footnote{\url{www.projectclearwater.org}, last accessed May 2021} as the use-case for a Cloud based virtualized NFV application--- it has been widely used in research as a standard test-bed setup for NFV related work \cite{di_mauro_ip_2019, jalodia_deep_2019, jalodia_deep_2021, jalodia_residual_2022, cherrared_lumen_2018, mijumbi_topology-aware_2017, bendriss_forecasting_2017}. It consists of 6 main components (VNFCs)--- Bono, Sprout, Homestead, Homer, Ellis, and Ralf. A high level view of these VNFCs and their functionalities replicating a standard IP Multimedia Subsystem (IMS) architecture is as shown in FIGURE \ref{fig:clearwater}.

\subsection{Clearwater Service Level Objectives (SLO)}
We use raw network telemetry data and system monitoring metrics obtained via a standard realization of the Clearwater test-bed setup to define the SLIs and SLOs governing an informal SLA. We utilise these metrics as the foundations for the SLIs, which when matched with a target threshold or range form SLOs. These metrics were collected on a 30 second sampling frequency through Monasca\footnote{\url{www.monasca.io}, last accessed May 2021}, an open-source Python based monitoring service running on each of the Clearwater VNFCs. The list of these collected metrics is presented in Table \ref{T1}, and further details regarding the data is elaborated upon in Section \ref{experimental-setup}.  \\

\begin{table*}[!h]
\caption{Raw metrics collected through Monasca during Clearwater vIMS application monitoring. This data is available for each of the VNFCs, and is sampled every 30 seconds.}   
\label{T1}
\centering
\adjustbox{width=\linewidth}
{
\begin{tabular}{c|c|c|c|}
\cline{3-4}
\multicolumn{2}{c|}{} &
  \textbf{Metric Name} &
  \textbf{Semantics} \\ \hline
\multicolumn{2}{|c|}{\multirow{3}{*}{\textbf{CPU}}} &
  cpu.idle\_perc &
  Percentage of time the CPU is idle when no IO requests are in progress \\ \cline{3-4} 
\multicolumn{2}{|c|}{} &
  cpu.system\_perc &
  Percentage of time the CPU is used at the system level \\ \cline{3-4} 
\multicolumn{2}{|c|}{} &
  cpu.wait\_perc &
  Percentage of time the CPU is idle AND there is at least one IO request in progress \\ \hline
\multicolumn{1}{|c|}{\multirow{8}{*}{\textbf{Disk}}} &
  \multirow{2}{*}{\textbf{Disk}} &
  disk.inode\_used\_perc &
  The percentage of inodes that are used on a device \\ \cline{3-4} 
\multicolumn{1}{|c|}{} &
   &
  disk.space\_used\_perc &
  The percentage of disk space that is being used on a device \\ \cline{2-4} 
\multicolumn{1}{|c|}{} &
  \multirow{3}{*}{\textbf{IO Read}} &
  io.read\_kbytes\_sec &
  Kbytes/sec read by an IO device \\ \cline{3-4} 
\multicolumn{1}{|c|}{} &
   &
  io.read\_req\_sec &
  Number of read requests/sec to an IO device \\ \cline{3-4} 
\multicolumn{1}{|c|}{} &
   &
  io.read\_time\_sec &
  Amount of read time in seconds to an IO device \\ \cline{2-4} 
\multicolumn{1}{|c|}{} &
  \multirow{3}{*}{\textbf{IO Write}} &
  io.write\_kbytes\_sec &
  Kbytes/sec written by an IO device \\ \cline{3-4} 
\multicolumn{1}{|c|}{} &
   &
  io.write\_req\_sec &
  Number of write requests/sec to an IO device \\ \cline{3-4} 
\multicolumn{1}{|c|}{} &
   &
  io.write\_time\_sec &
  Amount of write time in seconds to an IO device \\ \hline
\multicolumn{2}{|c|}{\multirow{3}{*}{\textbf{Load}}} &
  load.avg\_1\_min &
  The normalized (by number of logical cores) average system load over a 1 minute period \\ \cline{3-4} 
\multicolumn{2}{|c|}{} &
  load.avg\_15\_min &
  The normalized (by number of logical cores) average system load over a 15 minute period \\ \cline{3-4} 
\multicolumn{2}{|c|}{} &
  load.avg\_5\_min &
  The normalized (by number of logical cores) average system load over a 5 minute period \\ \hline
\multicolumn{2}{|c|}{\multirow{3}{*}{\textbf{Memory}}} &
  mem.free\_mb &
  Mbytes of free memory \\ \cline{3-4} 
\multicolumn{2}{|c|}{} &
  mem.usable\_mb &
  Total Mbytes of usable memory \\ \cline{3-4} 
\multicolumn{2}{|c|}{} &
  mem.usable\_perc &
  Percentage of total memory that is usable \\ \hline
\multicolumn{1}{|c|}{\multirow{4}{*}{\textbf{Network}}} &
  \multirow{2}{*}{\textbf{In}} &
  net.in\_bytes\_sec &
  Number of network bytes received per second \\ \cline{3-4} 
\multicolumn{1}{|c|}{} &
   &
  net.in\_packets\_sec &
  Number of network packets received per second \\ \cline{2-4} 
\multicolumn{1}{|c|}{} &
  \multirow{2}{*}{\textbf{Out}} &
  net.out\_bytes\_sec &
  Number of network bytes sent per second \\ \cline{3-4} 
\multicolumn{1}{|c|}{} &
   &
  net.out\_packets\_sec &
  Number of network packets sent per second \\ \hline
\end{tabular}
}
\end{table*}

In our previous work \cite{jalodia_deep_2021}, we fulfill the gap of a lack of realistic SLO definitions used within proposed models in research, and define Clearwater SLOs by using a combination of over two SLIs while drafting each SLO rule \cite{bendriss_forecasting_2017}. To highlight the varying reason behind the loss of QoS at any time, we define four SLOs for the Clearwater VNF--- SLO\(_{1}\) (load), SLO\(_{2}\) (computation), SLO\(_{3}\) (disk), and SLO\(_{4}\) (input/output, or IO). In-depth elaboration on these SLO definitions has been provided in our previous work \cite{jalodia_deep_2021}.


Formally, let $\mathcal{L}$ denote the set of SLOs thus defined:
\begin{equation}
    \mathcal{L} = [SLO_1, SLO_2, SLO_3, SLO_4]
    \label{slo-set-equation}
\end{equation}
Equivalently:
\begin{equation}
    \mathcal{L} = [SLO_{load}, SLO_{computation}, SLO_{disk}, SLO_{io}]
    \label{slo-set-name-equation}
\end{equation}
The metrics captured by Monasca are at the granularity of the individual VNFCs as shown in FIGURE \ref{fig:clearwater}, and an SLO violation at any of the individual VNFCs triggers an SLO violation state for the Clearwater application service. Therefore, we ultimately define the SLOs at the application level, i.e. for the entire VNF as an application service. Thus, each data instance is associated with 4 SLOs as defined by $\mathcal{L}$ above, where $SLO_j, j \in |\mathcal{L}|$ assumes one of two states:
\begin{equation}
    SLO_j=
    \begin{cases}
      1, & \text{if}\ Violated \ \text{(at any VNFC)} \\
      0, & \text{otherwise}\ 
    \end{cases}
    \label{slo-states-equation}
\end{equation}



\begin{algorithm}[] 

\SetAlgoLined
\BlankLine

\textbullet \ \textbf{\textsc{Pre-Processing}}

\SetKwInOut{A}{Input}
    \A {
        Data:\ $\mathcal{D}  \in \mathbb{R}^{d} $ \\
    }
    
\SetKwBlock{B}{procedure }{end}
    \B (\textsc{Data Splits} ){
       Split the data in train, validation, and test sets
    }
    
\SetKwBlock{C}{procedure}{end}
    \C (\textsc{Data Transformation}){
        Data standardization and normalization
    }

\SetKwBlock{D}{procedure}{end}
    \D (\textsc{Data Windowing and Batching}){
        Split data into sliding windows of features and targets\\
    }


\BlankLine

\textbullet \ \textbf{\textsc{GCRN Forecasting Model}} \\

\SetKwInOut{A}{Input}
    \A {
        Train and validation data windows from \textsc{Data Windowing}\\
    }
    
Define Clearwater VNFC graph $nodes$ \\
Initialize Clearwater graph $edge \ indices$ (connections) \\

\Repeat {Convergence}{
    \textbullet \ $X$ = Input data features sliding tensor slice \\
    \textbullet \ $Y$ = Expected data features sliding tensor slice \\
    \textbullet \ $\hat Y$ = Forecasted data features sliding tensor slice \\

    \textbullet \ GCRN Model Layers ($in, out, K$): \\
    GConvGRU layer (Clearwater $edge \ indices$) \ [\ref{gcrn-model}] \\
    Linear layer \\
    \textbullet \ Return: MSE loss between ($\hat Y, Y$) \\
}

\SetKwInOut{C}{Output}
    \C {
        Forecasted features $\hat Y$ \\
    }


\BlankLine

\textbullet \ \textbf{\textsc{Deep Reinforcement Learning Model}} \\

\SetKwInOut{A}{Input}
    \A {
        $x_t \in \mathcal{D}$ \\
    }

Initialize $Q$ network function with random weights $\theta$ \\
Initialize target $\hat Q$ function with random weights $\hat \theta$ \\
Initialize experience replay memory $\mathcal{E}$ to set capacity \\

\BlankLine

\Repeat {Training Episode Counter Complete}{
\textbf{\textsc{Training Deep Q-Network}} \\
    \textbullet \ Observe state $s_t$ (i.e. $x_t$) of Clearwater forecasting environment at time $t$ \\
    \textbullet \ Policy $\pi$: $\epsilon$-greedy \\
    \begin{algorithmic}[]
    \IF{Exploration (with probability $\epsilon$)}
    \STATE Randomly select action $a_t$ 
    \ELSE
    \STATE Select action $a_t = \text{argmax}_a Q(s_t,a;\theta)$
    \ENDIF
    \end{algorithmic}
    \textbullet \ Record reward $r_t$ for action $a_t$ \\
    \textbullet \ Store transition $(s_t, a_t, r_t, s_{t+1})$ in $\mathcal{E}$ \\
    \textbullet \ Sample random minibatch of transitions $(s_i, a_i, r_i, s_{i+1})$ from $\mathcal{E}$ \\
    \textbullet \ Perform a gradient descent step on $(r_i-Q(s_i, a_i;\theta))^2$ with respect to $\theta$ \\
    \textbullet \ Reset $\hat Q = Q$ periodically\\
}

\SetKwInOut{C}{Output}
    \C {
        Trained reinforcement learning model \\
    }

\caption{A GNN-based Framework for Topology-Aware Proactive SLA Management in a Latency Critical NFV Application Use-case}
\label{algorithm-1}

\end{algorithm}




\section{Topology-Aware Proactive SLA Management Framework} \label{proactive-sla-framework}
Given that the aim is to be able to proactively mitigate the potential SLA violations given the time-series of tracked system and application metrics at a set sampling frequency, we decompose the problem as that of continuous feature forecasting, followed by a reinforcement learning methodology that oversees the scaling policy to avoid potential SLA violations based on the forecasts.

\subsection{Graph Neural Network (GNN) based Clearwater Feature Forecasting}
Recurrent neural networks (RNN) are a class of neural networks that are powerful for modeling sequential data such as time-series, and are especially crafted towards such use-cases \cite{jalodia_deep_2021}. An RNN layer maintains an internal state that encodes information about the time-steps it has seen so far. However, traditional forecasting by way of evolved RNN models like GRU and LSTM consider only temporal information of features for sequence modeling.

When considering use-cases with additional spatial dependencies between features, relying solely on temporal variation likely imposes a cap on performance no matter how complex the RNN model in use. When input data can inherently be represented in a graph-based format of nodes and edges, the temporal flow of information between nodes also involves spatial dependencies. GNNs are a family of neural networks that deal with signals defined over graphs. As such, Graph Convolutional Recurrent Network (GCRN) is an evolved version of GNN that is a generalisation of classical RNN to data structured in a graph format, and can be defined as a deep learning model capable of predicting structured sequences of data. For computational efficiency, GCRN combines Convolutional Neural Networks (CNN) on graphs to identify spatial structures, and RNN to find dynamic patterns \cite{seo_structured_2016}. 

\begin{figure}[t]
	\centering
	\includegraphics[width=\linewidth]{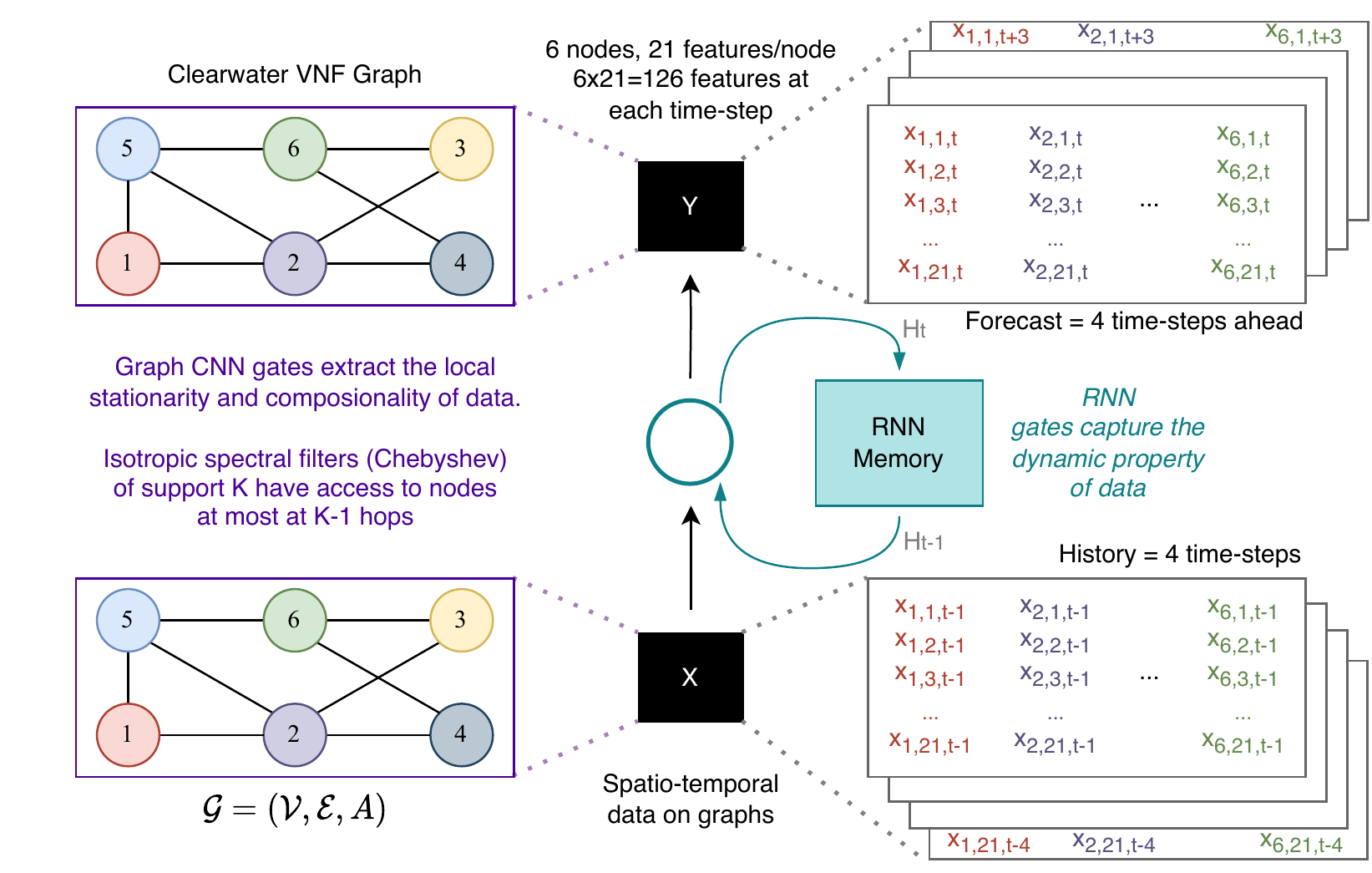}
	\caption{Overview of the GCRN-based Clearwater feature forecasting model.}
	\label{fig:gcrn-framework}
\end{figure}


Consider $x_t \in \mathcal{D}$ to be an observation at time $t$, where $\mathcal{D}$ denotes the domain of observed features. The Clearwater vIMS architecture as defined in FIGURE \ref{fig:clearwater} can be defined as an undirected and unweighted graph $\mathcal{G}=(\mathcal{V},\mathcal{E},A)$ as shown in FIGURE \ref{fig:gcrn-framework}, where $\mathcal{V}$ is a finite set of $|\mathcal{V}|=n=6$ vertices denoting the 6 Clearwater VNFCs, $\mathcal{E}$ as the set of edges, and $A \in \mathbb{R}^{n \times n}$ as the adjacency matrix denoting the (optional) weight of connection between two vertices. Therefore, a signal $x_t : \mathcal{V} \longrightarrow \mathbb{R}^{d_x}$ defined on the nodes of the Clearwater graph may be regarded as a matrix $x_t \in \mathbb{R}^{n \times {d_x}}$, whose column $i$ is the $d_x$-dimensional value of $x_t$ at the $i^{\text{th}}$ node.

FIGURE \ref{fig:gcrn-framework} provides an overview of the developed forecasting model. We define this Clearwater GCRN model with GRU as the RNN component, i.e. as a Chebyshev Graph Convolutional Gated Recurrent Unit Cell (GConvGRU) with Chebyshev filter \cite{seo_structured_2016} of size $K$. $K$ here controls the communication overhead, i.e. the number of nodes a given node $i$ should exchange information with in order to compute its local states. 

Mathematically, we model the forecasting problem such that we take a history of each of the Clearwater VNFC features for the past $H=4$ time steps, and forecast the evolution of these features over the next $F=4$ time steps:

\begin{equation}
\begin{gathered}
    \hat Y_{t},...,\hat Y_{t+F} = \\
    argmax_{Y_{t},...,Y_{t+F}} P(Y_{t},...,Y_{t+F} | X_{t-H},...,X_{t-1})
    \label{sequence-modelling}
\end{gathered}
\end{equation}

$X_t \in \mathcal{D}$ here is the matrix of the numeric state of each of the 21 features for the 6 VNFCs as observed at time $t$, $Y_t \in \mathcal{D}$ denotes the corresponding matrix of expected data features, and $\hat Y_t \in \mathcal{D}$ denotes the corresponding matrix of forecasted data features. $P$ models the probability of expected features $Y$ in a window of size $F$ to appear conditioned on the past $X$ features in a window of size $H$. Since we have spatial dependencies, the features of observations within $X$, $Y$, and $\hat Y$ are linked by pairwise relationships, modelled by $\mathcal{G}$.

Since we use a GRU based GCRN, the model (GConvGRU) can be represented as:
\begin{equation}
\begin{gathered}
    z = \sigma(W_{xz \ ^*\mathcal{G}} \ X_t + W_{hz \ ^*\mathcal{G}} \ h_{t-1}), \\
    r = \sigma(W_{xr \ ^*\mathcal{G}} \ X_t + W_{hr \ ^*\mathcal{G}} \ h_{t-1}), \\
    \tilde h = tanh(W_{xh \ ^*\mathcal{G}} \ X_t + W_{hh \ ^*\mathcal{G}} \ (r \odot h_{t-1})) \\
    h_t = z \odot h_{t-1} + (1-z) \odot \tilde h
    \label{gcrn-model}
\end{gathered}
\end{equation}

$z$, $r$, $\tilde h$, and $h_t$ are GRU parameters---  $z$ represents the update gate, $r$ represents the reset gate, $h_t$ represents the hidden state at time $t$, and $\tilde h$ represents the new hidden state. Further, $\odot$ represents the Hadamard product, and $\sigma$ represents the logistic sigmoid function. ${^*\mathcal{G}}$ then represents the graph convolution operator, and the support $K$ of the graph convolutional kernels defined by the the Chebyshev coefficients $W_{x\cdot}$ and $W_{h\cdot}$ determines the number of parameters independent of the number of nodes $n$.

Algorithm \ref{algorithm-1} presents the procedural workflow adopted towards achieving such a model that learns spatio-temporal structures from graph-structured and time-varying data. We evaluate this methodology in the following sections, benchmarking it against the current state-of-art on the use-case.

\subsection{Deep Reinforcement Learning (DRL)}
When it comes to dynamic application scenarios, RL is an effective machine learning methodology. A RL agent directly interacts with the environment to form a policy for decision-making based on a reward mechanism, that is customizable to achieve the desired outcomes. Specifically, RL follows the Markov Decision Process (MDP) model to train an agent that iteratively observes the state $s_t$ of the environment at a discrete time step $t$ to prescribe action $a_t$ that maximizes the reward $r_t$. Cumulating and maximizing the expectation of rewards over time, RL thereby attains an efficient policy $\pi$ for stochastic scenarios. Q-Learning in this regard is a typical off-policy model-free RL algorithm that calculates the value of each state-action pair as a Q-value function $Q(s,a)$:

\begin{equation}
\begin{gathered}
    Q(s_{t+1},a_{t+1}) = 
    \mathbb{E}\big[ r_t + \gamma \ max_{a'} Q_i(s',a') | (s_t,a_t)\big] \\
\end{gathered}
\end{equation}

Then, based on the policy $\pi = P(a|s)$ (e.g. $\epsilon$-greedy), it chooses the action with the largest Q-value ($Q^*$), and follows the gradient towards higher rewards:

\begin{equation}
\begin{gathered}
    Q^*(s,a) = \\
    max_{\pi} \mathbb{E}\big[ r_t + \gamma \ r_{t+1} + \gamma^2 r_{t+2}\cdots | s_t=s, a_t=a, \pi \big] \\
\end{gathered}
\end{equation}

$\gamma$ here is the discount factor applied to the rewards at each time step, accounting for the diminishing value of the reward at time $t$ as we iterate forward. DRL uses deep neural networks as non-linear function approximators to estimate the action-value function and deal with a large range of outcomes and their impact over time:

\begin{equation}
    Q(s,a;\theta) \approx Q^*(s,a)
\end{equation}

$\theta$ here refers to the weights of the neural network function approximator. Further, with the use of experience replay and network cloning, DRL is adept to deal with a complex environment with a big range of outcomes \cite{mnihHumanlevelControlDeep2015}.

\begin{figure}[t]
	\centering
	\includegraphics[width=\linewidth]{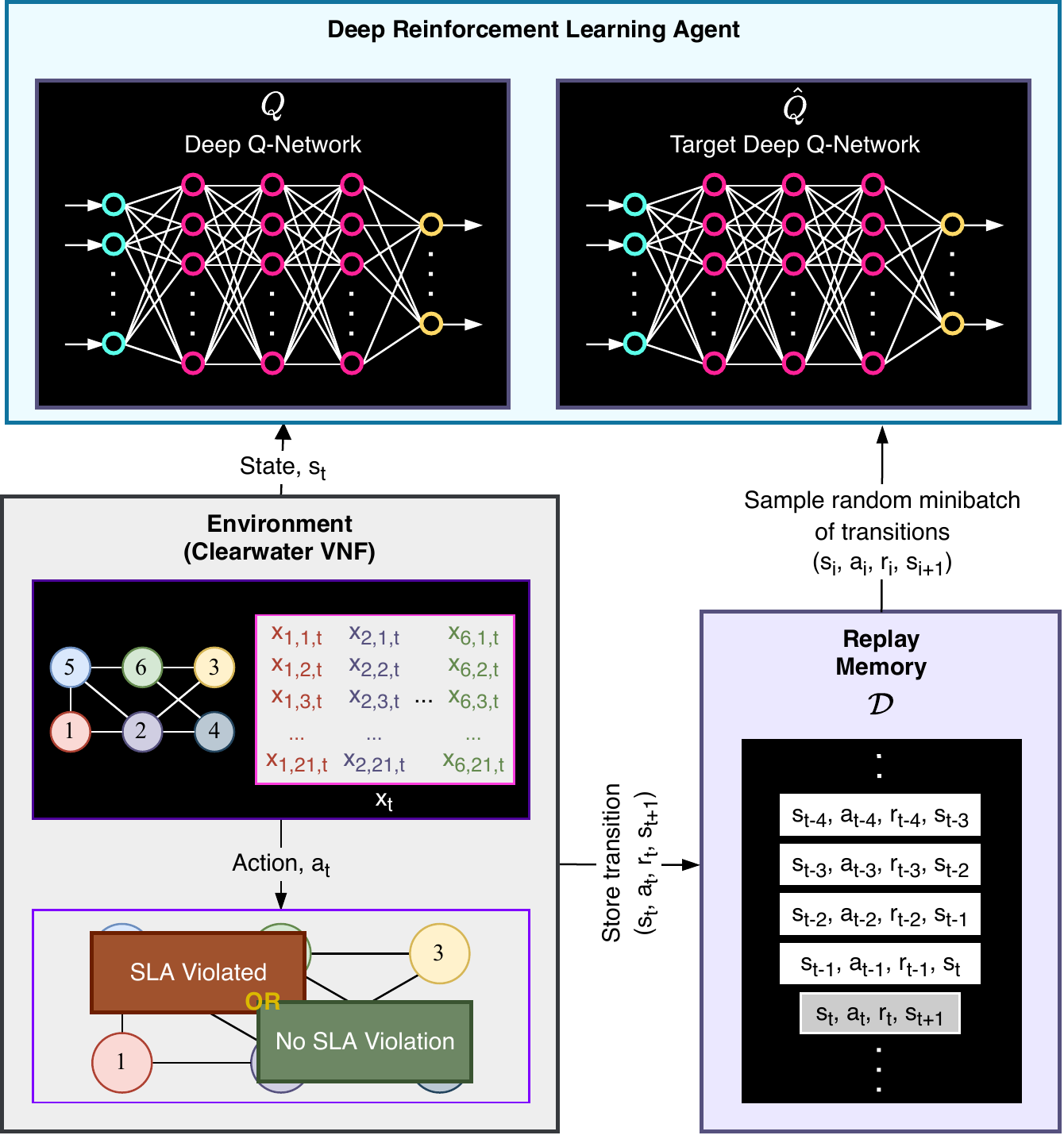}
	\caption{Overview of the deep Q-learning based Clearwater DRL model.}
	\label{fig:drl-framework}
\end{figure}

As stated previously, there is an active trade-off between efficiency and reliability in the existing scaling policies for latency-critical applications by service providers \cite{mijumbi_topology-aware_2017}. The aim of the reinforcement learning module here is to oversee and interface between an existing scaling policy and the defined SLA, and proactively generate warning alerts for anticipated SLA violations based on the GNN-based feature forecasting module and the defined SLOs. Building upon our existing work on granular SLA and SLO violation prediction\cite{jalodia_deep_2021}, this serves as a proof-of-concept for dynamic SLA-aware policy adjustment.

For the deep Q-learning problem formulation, we consider the SLA violated (state 1) if any of the 4 SLOs as defined in Section \ref{defining-sla-slo} are violated, i.e.,

\begin{equation}
    SLA=
    \begin{cases}
      1, & \text{if}\ \text{any} \ SLO_j \ \text{in} \ \mathcal{L} \ \text{is} \ Violated \\
      0, & \text{otherwise}\ 
    \end{cases}
    \label{sla-equation}
\end{equation}

Considering the information collected in the data we're working with, the action space is defined as:
\begin{equation}
    a_t=
    \begin{cases}
      1, & \text{i.e.}\ Scale \ up \\
      0, & \text{i.e.}\ No \ requirement \ to \ scale \ up
    \end{cases}
    \label{action-equation}
\end{equation}

Correspondingly, the reward function is defined as:

\begin{equation}
    r_t=
    \begin{cases}
      +20, & \text{if}\ a_t=1 \ \text{and} \ SLA=1 \\
      -10, & \text{if}\ a_t=0 \ \text{and} \ SLA=1 \\
      -5, & \text{if}\ a_t=1 \ \text{and} \ SLA=0 \\
      0, & \text{otherwise}\ 
    \end{cases}
    \label{reward-equation}
\end{equation}

Algorithm \ref{algorithm-1} details the workflow of the reinforcement learning model within the overall framework deployed, and FIGURE \ref{fig:drl-framework} presents its corresponding graphical overview. 

\begin{figure}[t]
	\centering
	\includegraphics[width=\linewidth]{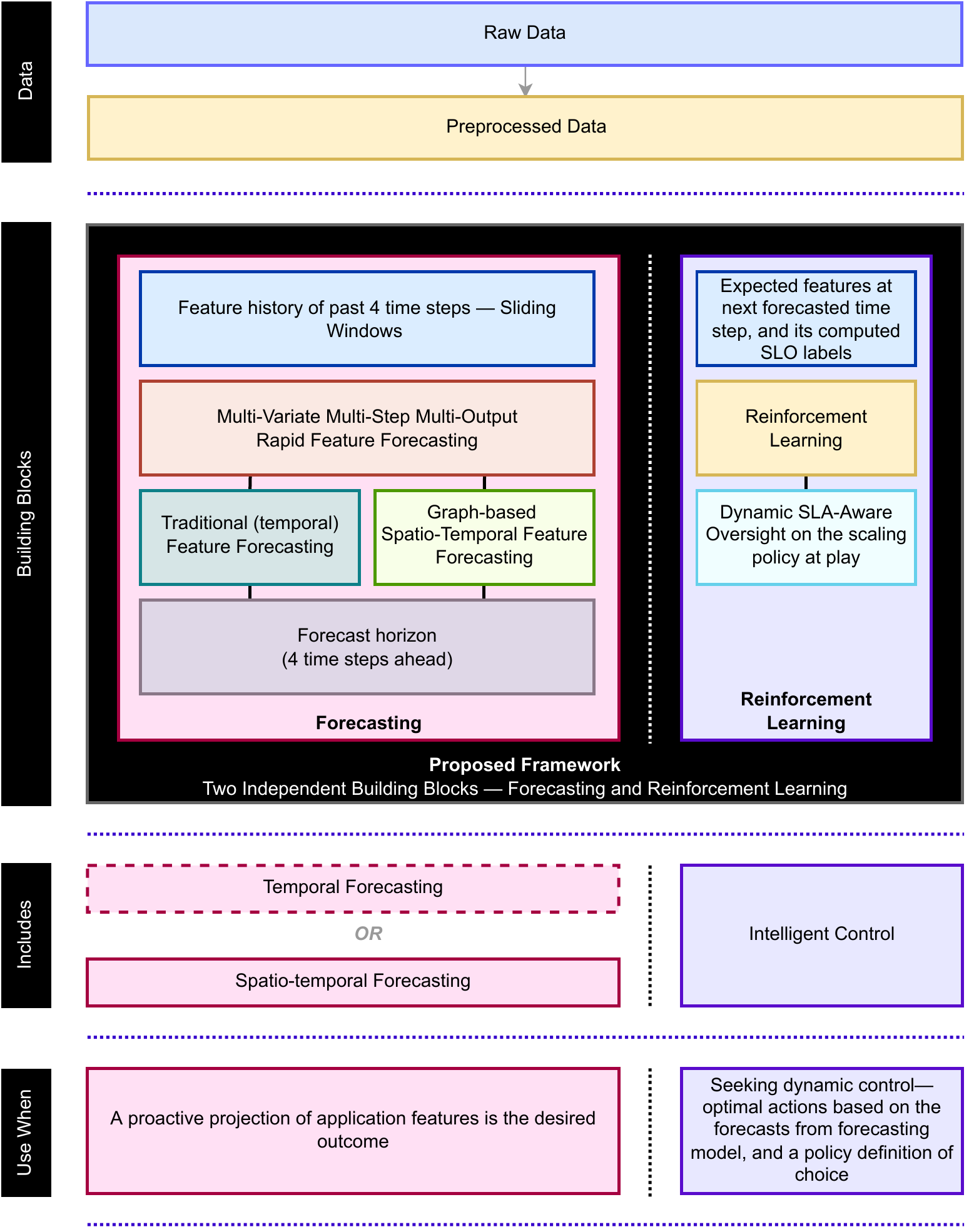}
	\caption{A high-level overview of the framework's building blocks, and what they deliver.}
	\label{fig:framework-blocks}
\end{figure}

FIGURE \ref{fig:framework-blocks} provides another view at how GCRN and GRL work together within the framework, and what they deliver independently. Further, the following sections elaborate on the learning and adaptation for the developed framework.

\section{Experimental Setup} \label{experimental-setup}
The experiments were all set up using Python (version 3.9.7) and its associated data-science libraries. We use PyTorch\footnote{\url{www.pytorch.org}, last accessed March 2022} and PyTorch Lightning \footnote{\url{www.pytorchlightning.ai}, last accessed March 2022} to program all the deep learning based implementations in the paper. For the GCRN implementation, we use PyTorch Geometric Temporal \cite{rozemberczki_pytorch_2021}, a temporal GNN extension library for PyTorch Geometric that supports the processing of spatio-temporal signals. We program a custom Gym \footnote{\url{www.gym.openai.com}, last accessed March 2022} environment that supports Clearwater data for the DRL implementation, and use Stable Baselines3 \footnote{\url{www.github.com/DLR-RM/stable-baselines3}, last accessed March 2022} to implement deep Q-learning in PyTorch.

\subsection{Dataset}
We use a publicly hosted dataset\footnote{\url{https://bit.ly/3gPY8c5}, last accessed May 2021} obtained via a standard Clearwater testbed, a visualization for which is presented in FIGURE \ref{fig:clearwater-testbed}. The dataset comprises of raw telemetric data files that track 26 metrics for each of the 6 monitored VNFs that compose the Clearwater ecosystem, and includes bursts of abnormal behaviour through its integrated stress testing tools to simulate VNF congestions and QoS degradations. The data is sampled every 30 seconds, and spans an overall period of 2 months. This corresponds to 156 features overall, and 177,098 rows of raw temporal data. A brief description of the captured metrics is summarised in Table \ref{T1}.

\subsection{System Configuration}
The experiments were performed in a Docker\footnote{\url{www.docker.com}, last accessed March 2022} based containerized environment running atop a bare-metal Linux server with 2 NVIDIA\textsuperscript{\textregistered} Tesla K20m GPUs. The Docker image runs an Ubuntu 20.04 LTS operating system, and CUDA version 10.2 for the GPUs. 

\subsection{Learning and Adaptation}
We adopt a sliding window methodology for time-series forecasting, which is suitable for rapid forecasting. The models were tested on different window sizes, and considering the short-term dynamics of the use-case, the input window size was optimally set to 4 time-steps (i.e. data from the past 2 minutes, with a 30 second sampling frequency), and the models forecast the possibilities of SLO violations over each of the next 4 time-steps (i.e. the future 2 minute horizon). Thus, we use the data for all the features over the last 2 minutes to forecast the specific SLOs that may be violated at each step over the next 2 minutes. 


The available data consisting of 177,098 rows was split into training and test sets in the ratio of $80:20$, with the training set then further split into training and validation set in same $80:20$ ratio split. Further, given the high degree of extreme outliers in data due to the incorporated stress tests, and the widely varied scales for each of the features depending on the category of raw metric, we use a Quantile Transformer as a non-linear transformation method to transform the very skewed nature of this dataset and map it to a uniform distribution in $[0,1]$. 

For training the forecasting model, we use Mean Squared Error (MSE) as the loss function to be minimized. Further, we use Adam as the optimizer for its adaptive nature, with a learning rate of $10^{-2}$. ReLu (Rectified Linear Unit) is used as the activation function between layers due to its computational simplicity and high optimization performance \cite{goodfellow_deep_2016}. For the general layers, we use the default PyTorch initializations for weights, activation, and bias. To control overfitting as the model gains complexity, we deploy an early stopping criteria that tracks the validation MSE with a patience of 10 epochs to ensure that the training is not stopped at a local optimization minima. Model weights are restored from the best epoch at the end of training, i.e. the best performance achieved before the model began to overfit on the training set. \\

For the deep Q-learning model, the observation space is defined as a discrete snapshot of the forecasted features at time $t$, i.e., the episode length is defined as 1. We use a multi-layer perceptron based deep-learning policy for the Q-value approximation and the target network. We train the model for 5,000,000 timesteps, set the learning rate to 0.0001, and stick to default values from original algorithm propositions in \cite{mnihHumanlevelControlDeep2015} and \cite{mnih_playing_2013} for most of the hyperparameters. To elaborate, the size of the replay buffer was set to 1,000,000, the minibatch size for each gradient update was set at 32, the update coefficient $\tau$ was set to hard update at 1, the discount factor $\gamma$ was set to 0.99. The model collects transitions for 50,000 timesteps before learning starts, and the model is set to update every 4 steps, with 1 gradient step after each rollout. The replay buffer class was set to automatic selection, and the target network $\hat Q$ was set to update every 10,000 environment steps, with the maximum value for the gradient clipping was set to 10. The initial value of the random action probability $\epsilon$ was set to 1, while the final value of $\epsilon$ was set to 0.05, with the exploration fraction set to 10\% of the training period.

\begin{table}[]
\caption{Performance of two GCRN model architectures on forecasting metrics.}
\label{T2}
\centering
\adjustbox{width=\linewidth}
{
\begin{tabular}{|c|c|c|c|c|c|c|}
\hline
\textbf{Model} & \textbf{\begin{tabular}[c]{@{}c@{}}In\\ Channels\end{tabular}} & \textbf{\begin{tabular}[c]{@{}c@{}}Out\\ Channels\end{tabular}} & \textbf{\begin{tabular}[c]{@{}c@{}}Filter\\ Size\end{tabular}} & \textbf{MSE}   & \textbf{MAE}   & \textbf{RMSE}  \\ \hline
GCRN           & 4                                                              & 128                                                             & K = 3                                                          & 0.036          & 0.132          & 0.188          \\ \hline
GCRN           & 4                                                              & 2048                                                            & K = 3                                                          & \textbf{0.034} & \textbf{0.128} & \textbf{0.185} \\ \hline
\end{tabular}
}
\end{table}

\begin{figure}[t]
	\centering
	\includegraphics[width=\linewidth]{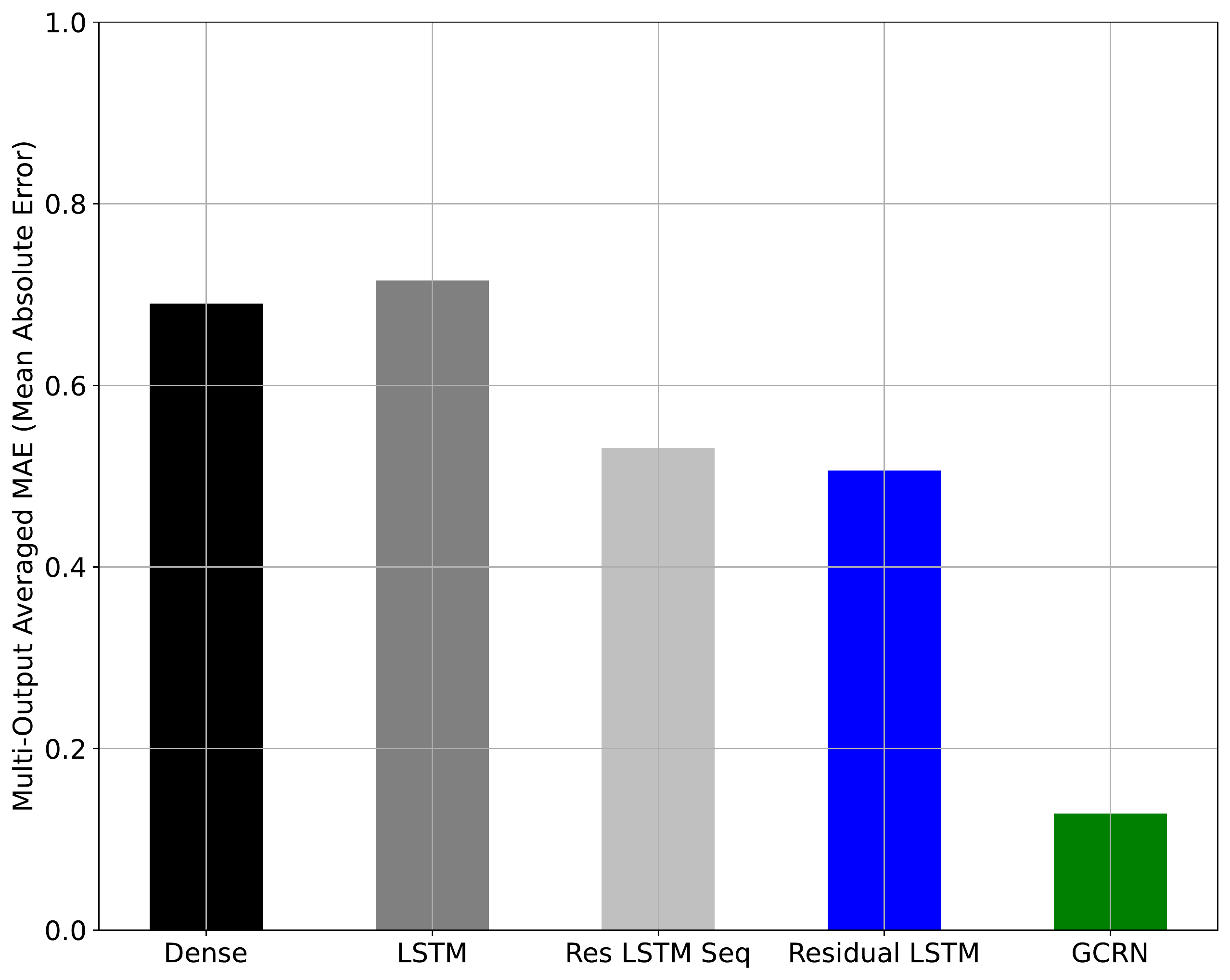}
	\caption{Multi-output forecasting performance (MAE) of proposed Clearwater GCRN model against the previously established state-of-art models on the use-case, all configured with the same 2048-dimensional wide model architecture.}
	\label{fig:forecast_samearch}
\end{figure}

\section{Results and Discussion} \label{results-discussion}
Table \ref{T2} presents the results for the performance of two architecture variants of the GCRN model on forecasting metrics. To set this in perspective, FIGURE \ref{fig:forecast_samearch} compares the mean absolute error (MAE) of various models that have been previously applied to the use-case data \cite{jalodia_residual_2022} with the same forecasting horizon. Our proposed GCRN-based spatio-temporal forecasting model achieves a 74.62\% improvement over the Residual-LSTM \cite{jalodia_residual_2022} model that was proposed as the best-in-class in previous work. This quantifies the significant leap in performance when considering spatio-temporal GCRN as opposed to a general feed-forward deep learning model (Dense), and specialized RNN-based deep learning models like LSTM and its Residual variants which have been covered in detail in \cite{jalodia_residual_2022}. GCRN merges CNN for graph-structured data and RNN to simultaneously identify meaningful spatial structures and dynamic patterns, and this achieves a significant leap in performance on the use-case as opposed to regular and temporal deep learning models.

\begin{figure}[t]
	\centering
	\includegraphics[width=\linewidth]{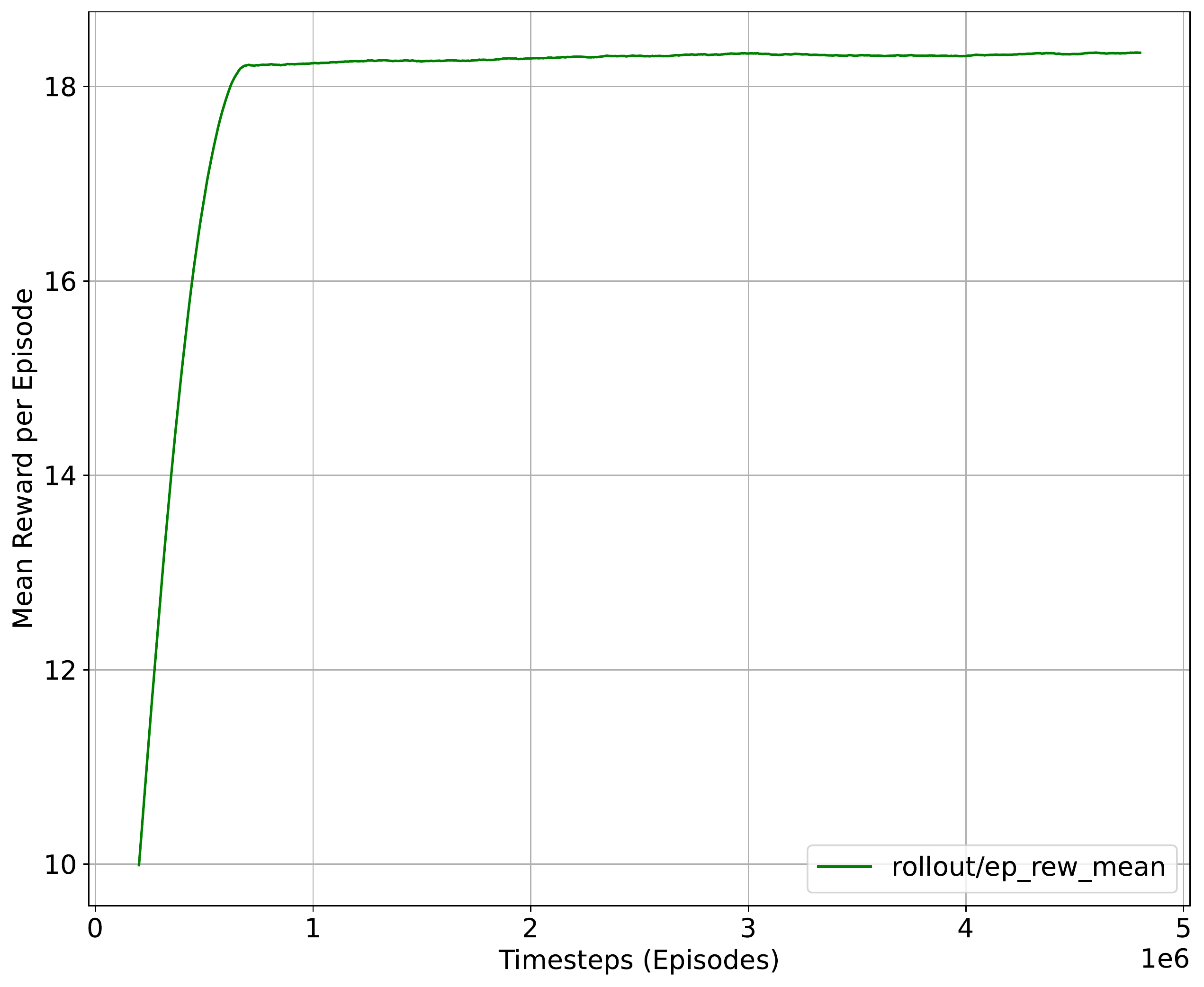}
	\caption{Mean episodic DRL training reward (averaged over 100 episodes, further smoothed with a rolling mean of 100,000 timesteps for a better visualization of trend).}
	\label{fig:training_reward}
\end{figure}

\begin{figure}[t]
	\centering
	\includegraphics[width=\linewidth]{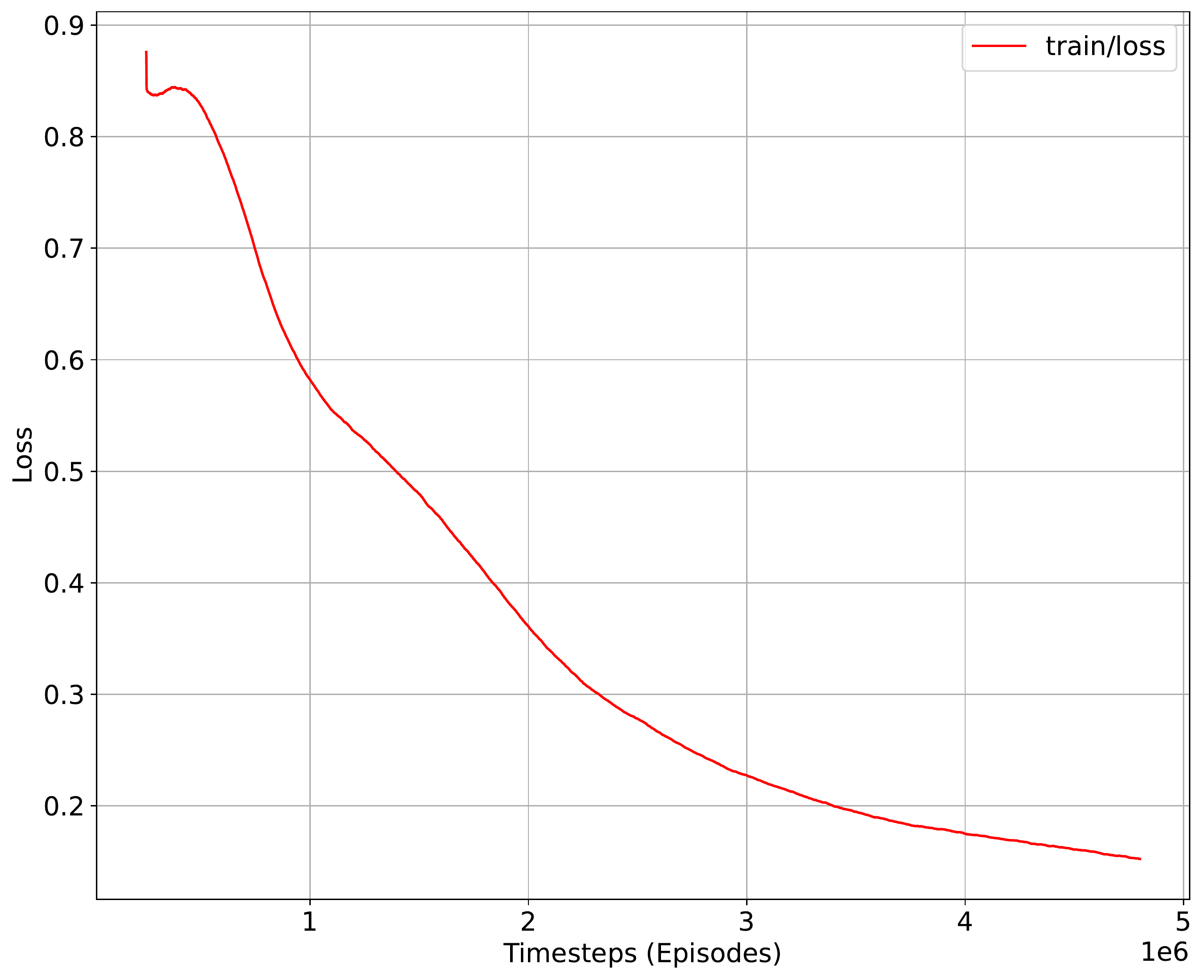}
	\caption{Total loss value observed at each timestep as DRL training progresses (smoothed with a rolling mean of 100,000 timesteps for a better visualization of trend).}
	\label{fig:training_loss}
\end{figure}

\begin{figure}[t]
	\centering
	\includegraphics[width=\linewidth]{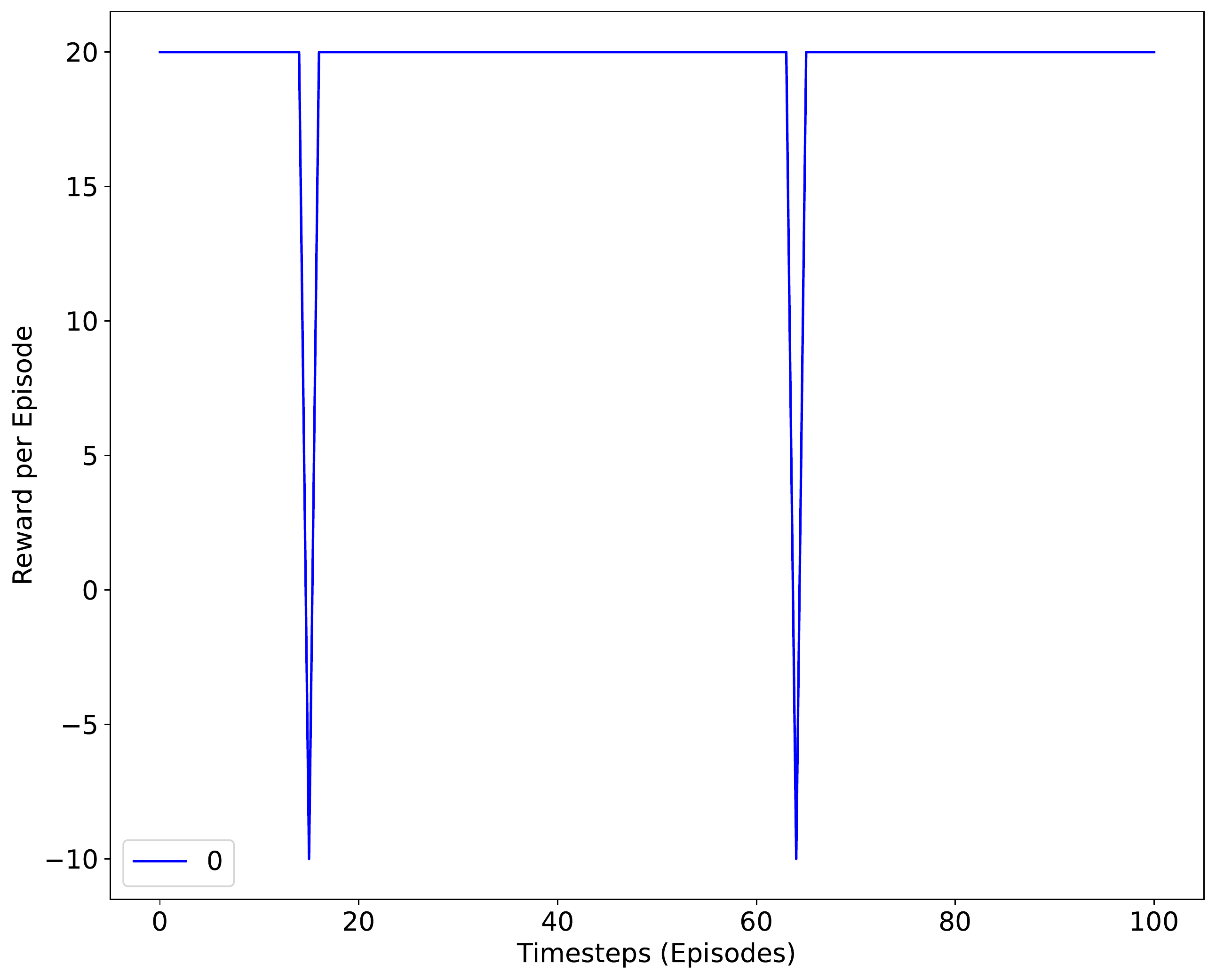}
	\caption{Reward observed per episode when testing the trained deep Q-learning model.}
	\label{fig:prediction_reward}
\end{figure}

Further, FIGURES \ref{fig:training_reward} and \ref{fig:training_loss} demonstrate good convergence of the proposed deep Q-learning model while training, with incremental improvements in learned policy. FIGURE \ref{fig:prediction_reward} shows the results when the trained model policy was tested on a random test sample of 100 episodes, delivering a positive reward 98\% of the time. 

\section{Conclusion, Challenges, and Future Work} \label{conclusion-futurework}
In this work, we propose a GNN and DRL based framework for proactive SLA management in the use-case of Clearwater, a latency-critical NFV application. We compose a  Graph Convolutional Recurrent Network (GCRN) based spatio-temporal multivariate time-series forecasting model that forecasts the evolution of system monitoring features for the Clearwater VNF over the next 4 time steps, delivering 74.62\% improved performance over the established baseline state-of-art model on the use-case. The wide leap in forecasting performance with our proposed framework quantifies the benefits of incorporating spatial metadata in use-cases that require multi-dimensional feature forecasting in high-frequency temporal data flows. Further, we leverage realistic Clearwater SLA and SLO definitions to develop a deep Q-learning based DRL model, and train it to act as an effective SLA-aware dynamic oversight of the scaling policy at play. \\

As methodologies and principles evolve across systems and domains, and new principles like site reliability engineering take over the legacy operational practices, machine learning has a key role to play. The next generation of systems contain requirements that move beyond traditional monitoring, now venturing into the domain of observability. Proactive problem and event management is the need of the hour to support Cloud-based infrastructure and services for next generation verticals. To this end, one of the key challenges in production systems is to tackle data drift--- models trained offline and then deployed online are effective only as long as the data distribution remains equivalent to what the model has been trained on, and the features' behaviour lies within the sample subset of scenarios that the training set contains. To tackle this, it is important to configure retraining feedback loops that periodically expose the model to a new subset of incoming data offline, allow a readjustment of hyperparameters inline with the new observations if any, and then a redeployment of the updated model in production. While this is beyond the scope of the current manuscript, the step has been highlighted in FIGURE \ref{fig:clearwater-testbed}. Moreover, an enhanced ability to react to the unknown is another reason why reinforcement learning is an important layer of oversight over the supervised learning models. \\

In future work, we plan to set up a custom testbed deployment with the aim to better mimic production systems and test the model overheads in retraining feedback loops, and improve the reinforcement learning model for extended online sequence modeling with longer episode lengths. Retraining overheads are a key consideration in production deployments, and something that could become a potential demerit of an otherwise well-performing model. Further, while the data we worked with includes bursts of abnormal behaviour through its integrated stress testing tools to simulate VNF congestions and QoS degradations, the proposed approaches from this dissertation could also be integrated with a traffic and workload forecasting methodology for a higher degree of detail in proactive violations' prediction, and then combined with a dynamic policy enforcement for a wider end-to-end management control loop. Moreover, since the methodology is transferable to other latency-critical verticals within the high-availability network slice, we aim to validate this on a different use-case while also collecting and incorporating external network features.

\bibliographystyle{ieeetr}
\bibliography{main.bib}


\vspace{-2cm}

\begin{IEEEbiography}[{\includegraphics[width=1in,height=1.25in,clip,keepaspectratio]{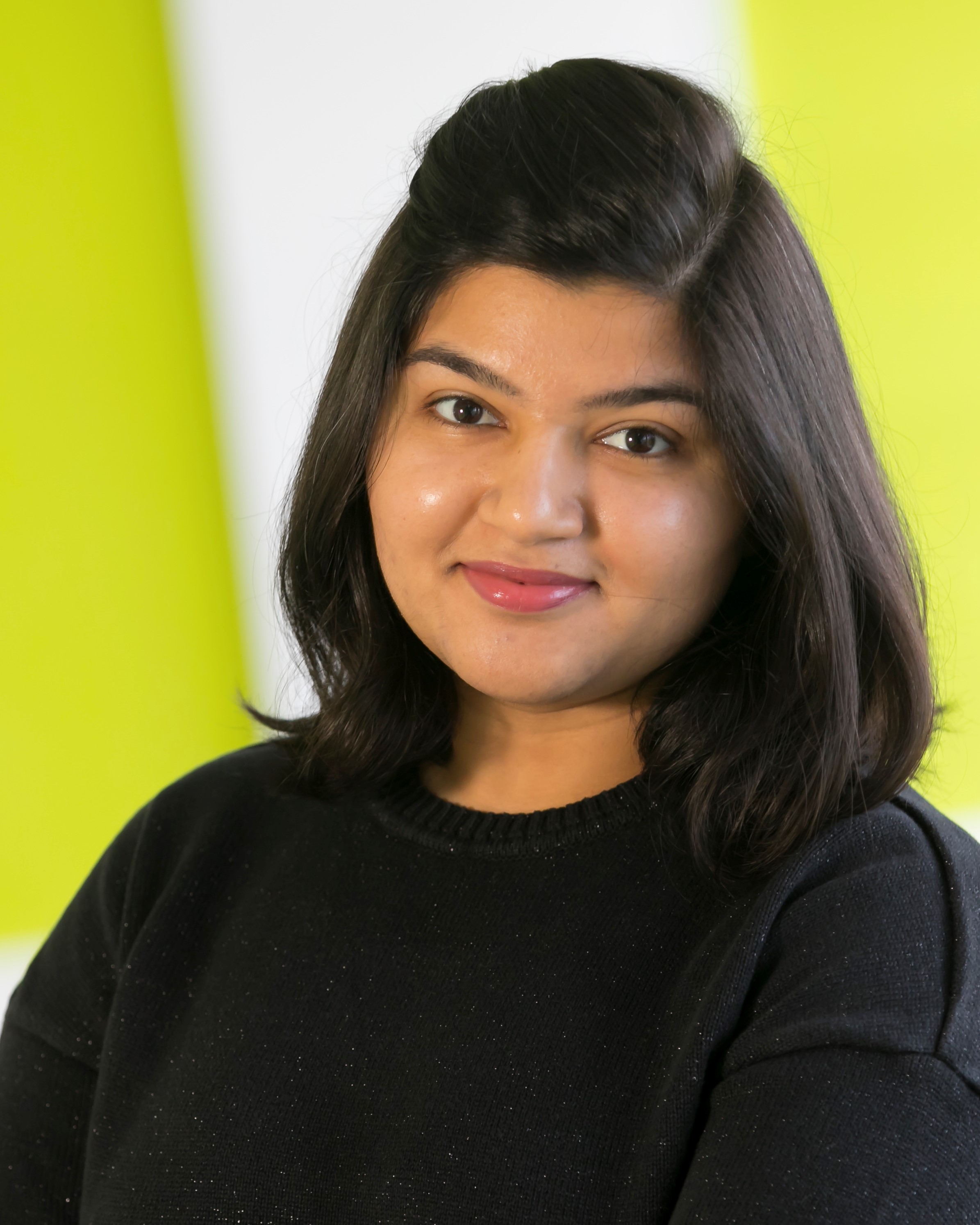}}]{Nikita Jalodia} recently completed her Ph.D. with the Department of Computing and Mathematics at the Emerging Networks Lab Research Division in Walton Institute of Information and Communication Systems Science, South East Technological University, Ireland. She joined in July 2017, and has since been working as a part of the Science Foundation Ireland funded CONNECT Research Centre for Future Networks and Communications. Her current research interests include Data Science, Network Function Virtualization (NFV), Machine and Deep Learning, Knowledge-Defined Networks, Internet of Things (IoT), and Fog and Cloud Computing. She received her Bachelor's Degree in Computer Science and Engineering from The LNM Institute of Information Technology, Jaipur, India in 2017, with a specialization in Big Data and Analytics with IBM. She has also previously worked as a Software Developer at Publicis (Sapient) Global Markets, India.
\end{IEEEbiography}%
\vspace{-2cm}
\begin{IEEEbiography}[{\includegraphics[width=1in,height=1.25in,clip,keepaspectratio]{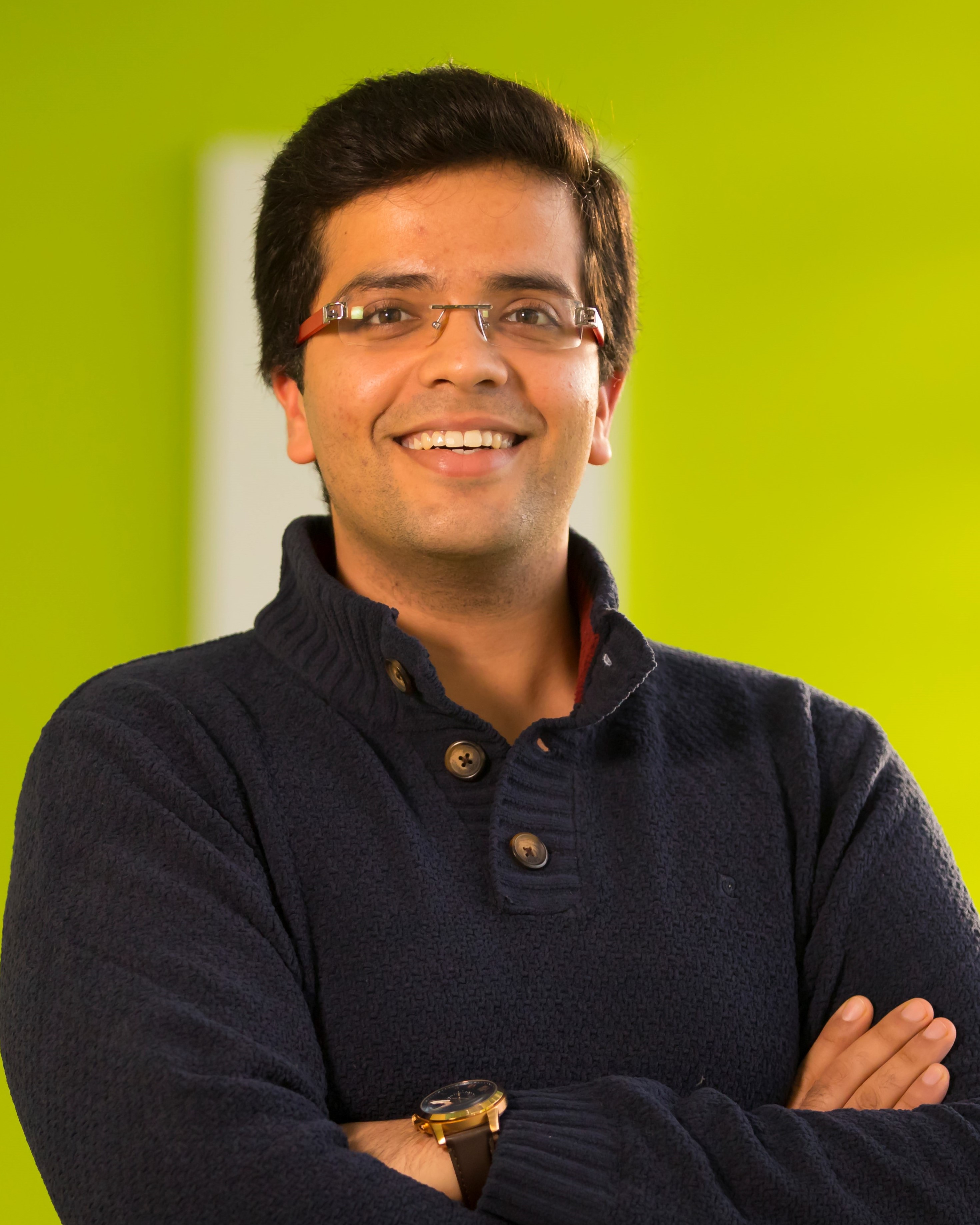}}]{Mohit Taneja} is a Tenured Lecturer in Business Information Systems, Data Science and Computing with the Department of Accountancy and Economics, School of Business at South East Technological University. Previously, he was a Postdoctoral Research Fellow with the Programmable Autonomous Systems Division of Walton Institute; working on EU, National, International, and Industry funded projects. He worked as the Project and Tech Lead on EU-H2020 funded Smart Cities 2030 project. In Walton, he also worked within the Strategic Division at Walton, liaising with all the Research Divisions within the Institute. He has also been associated with the SFI funded VistaMilk Research Centre and CONNECT Centre. He has previously worked as an Experienced Software Research Engineer with the Emerging Networks Lab (ENL) division. He earned his Ph.D. from South East Technological University, Ireland in 2020. The title of his dissertation was ‘Fog Computing Support for Internet of Things Applications’. He was also a visiting research fellow at IBM Research Labs, Ireland from 2017-2018. His current research interests include Fog and Cloud Computing, IoT, Distributed Systems, and Distributed Data Analytics. He received his Bachelor's Degree in Computer Science and Engineering from The LNM Institute of Information Technology, Jaipur, India in 2015.
\end{IEEEbiography}

\begin{IEEEbiography}[{\includegraphics[width=1in,height=1.25in,clip,keepaspectratio]{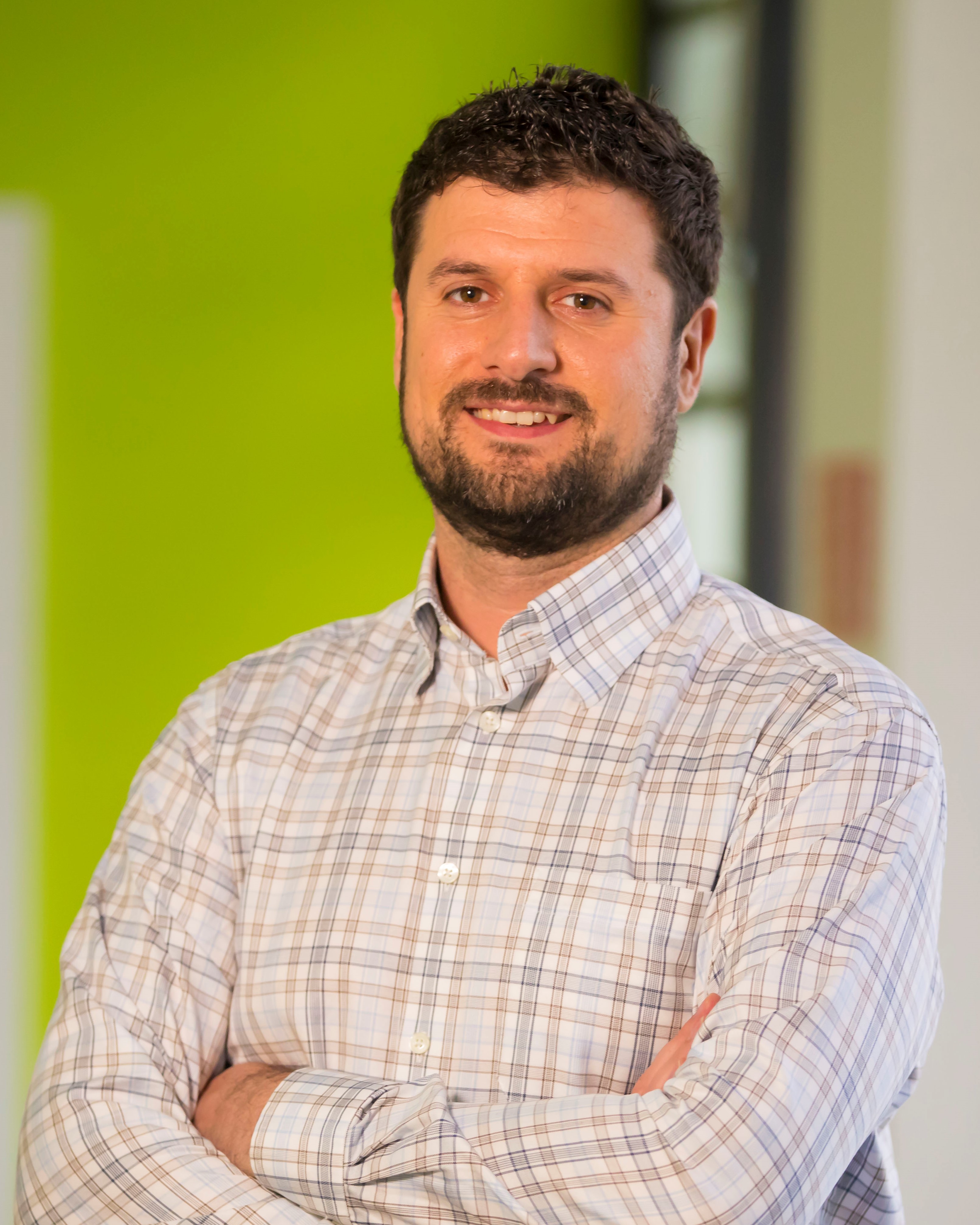}}]{Alan Davy} is currently the Head of the Department of Computing and Mathematics, School of Science and Computing, South East Technological University (SETU), Waterford, Ireland. Previously, he was the Research Division Manager of the Emerging Networks Laboratory in WIT's Walton Institute of Information and Communication Systems Science. He has been the Coordinator and Principal Investigator on a number of National, International, and EU projects such as TERAPOD, SFI-TIDA, 5GinFIRE's C2G-RAN, etc. He was awarded B.Sc. (with Hons.) in Applied Computing and Ph.D. degrees from the South East Technological University, Waterford, Ireland, in 2002 and 2008 respectively. He has been a recipient of the Marie Curie International Mobility Fellowship in 2010, and has worked at the Universitat Politècnica de Catalunya, and IIT Madras in India. His current research interests include Virtualised Telecom Networks, Fixed and Wireless Network Management, Software Defined Infrastructure, Internet of Things, Fog and Edge Computing,  Bio-inspired Systems, Molecular/Nano-scale Communications, and TeraHertz Communications.
\end{IEEEbiography}


\EOD
\end{document}